\documentclass[twocolumn,showpacs,preprintnumbers,amsmath,amssymb]{revtex4}
\usepackage{graphicx}
\usepackage{dcolumn}
\usepackage{bm}

\begin{document}

\title{Mechanisms for higher $T_C$ in copper oxide superconductors; Ideas from
band calculations.}

\author{T. Jarlborg}

\affiliation{
DPMC, University of Geneva, 24 Quai Ernest-Ansermet, CH-1211 Geneva 4,
Switzerland
}


\begin{abstract}
Band calculations for the hole doped La$_2$CuO$_4$ system show
that artificial periodicities of Ba dopants can give the material
different properties than from a uniform distribution of dopants.
A periodicity within the planes make static pseudogaps which
could be tuned to raise the density-of-states (DOS) at $E_F$ and the
superconducting $T_C$. 
 A periodic doping dependence perpendicular to the CuO planes
can increase the matrix element for spin fluctuations.

\end{abstract}


\maketitle


Band calculations for high-$T_C$ copper oxide systems have shown that phonon
distortions within the CuO-planes are favorable for anti-ferromagnetic (AFM) spin
fluctuations \cite{tj1}. Such type of spin-phonon coupling (SPC) can
be responsible for (dynamic) stripes and pseudogaps, since a periodic potential
$V_q exp(-i\vec{q}\cdot\vec{x})$ leads to at gap 2$V_q$ at the zone boundary \cite{tj3}. 
The qualitative results 
of ab-initio calculations and free-electron like parametrization are able to
describe several typical and unconventional features of the high-$T_C$'s \cite{tj5}.
For instance, the pseudogap is according to SPC more developed at low doping leading to a 
low density-of-states (DOS) at the Fermi energy ($E_F$) and is
therefore in competition with superconductivity 
 \cite{tj5}. The role of stripe order and the importance of non-uniform doping for the properties of the cuprates
has been discussed theoretically and experimentally \cite{mena,bozo2}.
There is no consensus about the mechanism of
superconducting pairing in the high-$T_C$ copper oxides, but a high DOS
at $E_F$, $N(E_F)$, is good for superconductivity in
theories based on electron-phonon coupling, $\lambda_{ep}$,
as well as on coupling due to spin fluctuations, $\lambda_{sf}$.  
 Here are proposed different ways of increasing $N(E_F)$ and/or the exchange interaction for
AFM spin fluctuations through periodic doping distributions. This is done through band calculations,
using the  Linear Muffin-Tin Orbital (LMTO)
method in the local spin-density approximation \cite{tj5},
for large unit cells with superstructures of La/Ba substitutions in La$_{(2-h)}$Ba$_h$CuO$_4$ (LBCO).
One calculation for a small cell is based on the virtual crystal approximation (VCA), 
where the nuclear and electronic La-charges (57.0) are reduced to (57-$h/2$) to account for
a perfectly delocalized doping, $h$, in holes per Cu. 

Two types of superstructures  
are considered. The first is for extensions of the unit cell parallel to the CuO planes along the
$\vec{x}$-axis.
The objective is to find
a mechanism to oppose the drop in $N(E_F)$ caused by the
pseudogap in underdoped LBCO, or even to increase $N(E_F)$ when there is no pseudogap
in overdoped cases.
The idea is to chose the strength ($V^*$) and wavelength ($\Lambda^*$) of a static potential modulation 
so that one of the resulting DOS peaks (above or below the induced gap) will be at $E_F$.

 A simple doubling of one La$_2$CuO$_4$ cell in the x-y-plane makes up the
basic AFM cell with opposite spin on the two Cu sites.
Large cells containing totally 56 or 112 sites are made by putting together four or eight of such AFM cells
along $\vec{x}$ (with lengths 4 or 8 times the lattice constant $a_0$), 
where Ba/La substitutions
create the potential modulations along the supercell. 
Three calculations are made for cells of length 8$a_0$. One is where the 4 La surrounding the two
Cu within the first row (along $\vec{y}$) are replaced with Ba.
The fraction of sites occupied by Ba, $f$, is 1 within this row, 
while no other Ba/La substitution is made in the other 7 rows further
down along $\vec{x}$. This makes the effective doping $h$ equal to 0.25 holes/Cu, and the 
"periodicity" ($\Lambda^*$) of the doping is 8$a_0$. A second case with
$h$=0.125, is considered in the same cell except that only
half of the La-sites are exchanged with Ba in zigzag-like pattern along $\vec{y}$ ($f=\frac{1}{2}$).
The same two sets of configurations are also made for a half as long cell along $\vec{x}$ 
("periodicity" 4$a_0$),
with doping $h$=0.5 and 0.25 respectively. The third case with 112 sites is with
two adjacent rows of complete La/Ba exchange, i.e. for $h=0.5$ and $f=2$.

\begin{table}[b]
\caption{\label{table1} 
Ab-initio LMTO results for periodic La/Ba substitution along $\vec{x}$.
The doping $h$ is in holes per Cu, and $f$ is the fraction of Ba occupation within the
doped rows ($f$=2 is for two adjacent rows of complete La/Ba exchange). 
The periodicity $\Lambda^*$ is the distance (in $a_0$-units) between the doped rows.
The  $N(E_F)$ (in states/Ry/Cu) is obtained from a rigid-band shift (to account for the doping, $h$) on 
the DOS of undoped La$_2$CuO$_4$, shown in Fig. 1. 
The largest charges ($Q$, electrons per Cu) and attractive
 potential shifts on Cu ($V^*$, mRy) are in regions far from the Ba-dopants.}
\vskip 5mm
\begin{center}
\begin{tabular}{l c c c c c }
\hline
 $h$  & $f$ & $\Lambda^*$ & $N(E_F)$ & $Q_{Cu} $ & $V^*$ \\
\hline \hline
 
 0.125 & $\frac{1}{2}$ & 8$a_0$ & 25.0 & 10.36$\pm.17$ & 6.6   \\
 0.250 & 1             & 8$a_0$ & 16.5 & 10.33$\pm.18$ & 11  \\
 0.250 & $\frac{1}{2}$ & 4$a_0$ & 16.5 & 10.36$\pm.11$ & 6.4 \\
 0.500 & 1             & 4$a_0$ & 10.0 & 10.31$\pm.15$ & 14.5 \\
 0.500 & 2             & 8$a_0$ & 10.0 & 10.31$\pm.23$ & 23.3 \\
\hline
\end{tabular}
\end{center}
\end{table}

The calculated variation of charge, $\Delta Q$, and potential shift, $V^*$, on Cu are shown 
in Table I. The charges on Cu near Ba-sites decrease because of an upward shift of the potential,
and the widths of the Cu-d band become narrower ($\sim$ 10-20 percent).
It is quite natural that $V^*$ is large for large $f$,
 since the influence on Cu will be larger with many Ba nearby.
The second trend, larger $V^*$ for short periodicities, is 
connected with large doping, lower $N(E_F)$
and less effective screening. The $V^*$ are of similar size (5-10 mRy) as
$V_q$ from typical phonon distortions \cite{tj6}.

The wave length $\Lambda$ (in units of $a_0$) of the phonon wave in SPC 
is proportional to the inverse doping, $\Lambda = 1/h$ 
(the wave length of the spin wave is twice as long). This is for not too large doping,
and $E_F$ is at the pseudogap \cite{tj5}.
The doping can be chosen so that $\Lambda^* \approx \Lambda$. 
But this is not the the main interest here, because even if
spin fluctuations are responsible for the pseudogap in SPC,
it is not obvious that a static gap should promote spin fluctuations.
 Instead, the goal is to
create humps in the DOS near $E_F$. 
In order to have a large $V^*$, and large humps
around a deep gap, it is probably good to have a sizeable La/Ba-exchange within the given row
and a clear separation between the rows, if the number
of dopants allow for that.

The imposed spacing $\Lambda^*$ between Ba-rich rows can be related to
a fictive doping $h^*$, $\Lambda^* = 1/(h+h^*)$, where $h^*$ is
number of states between $E_F$ and
the energy ($E^*$) at the static gap, with the requirement that $(E_F-E^*) \approx V^*$. 
From an effective
DOS between $E^*$ and $E_F$, $\Tilde{N}$, this defines a positive
$h^*= \Tilde{N} V^*$ (holes) for a positive $V^*$, with $E_F$ at the DOS-hump
above the static gap. This will be best
at underdoping, while the use of a hump below the gap (negative $V^*$ and $h^*$) is appropriate
in overdoped cases. 
The
fraction $f= h \Lambda^*$,
is not a free parameter, but it has to fit with the true doping.
A very small $f$ is not appealing, since the dopants will be diluted in many
closely spaced rows with very weak $V^*$. That would look more like
a random distribution of the dopants. Stronger modulations from well structured 
dopings are preferred.

Results for different dopings, shown in Table II, are based on interpolated values
from Table I.
For example, if the doping is 0.05 it is best to replace 1/3 of La with Ba within every 7th (6.7) row
for a good increase of the DOS at $E_F$. Higher doping $h=0.15$ would require
1/2 (0.54) La-Ba replacements within every 4th (3.6) row, and so on.
"New" gaps would appear 4-6 mRy below $E_F$ in these cases. A case with extreme overdoping, $h=0.48$,
can hardly support strong SPC because the wave would be too short. But if it should be made it would require
 complete La/Ba exchange within 2 rows ($f=2$) separated by $\sim 2$ rows
of pure La ($\Lambda^* \approx 4-5$) to increase the DOS at $E_F$. The static gap would be
 $\sim$20 mRy above $E_F$. The gap is wide because $V^*$ is large when all dopants are concentrated
within few layers, and it forces the modulation to be short. The last line in Table II, is a suggestion for
a sinusoidal profile of La/Ba-exchange 
along the cell ($f = 3.3$ but the doping is spread over more than 3 rows), in order to diminish the effective
$V^*$ and to have a more reasonable $\Lambda^*$.

It is difficult to quote precise enhancements of the DOS at $E_F$ for these periodic dopings.
But in previous calculations with comparable 
amplitudes of $V_q$ for phonon distortions, the enhancements can easily be 30-50 percent within peak widths of
the order 3 mRy \cite{tj5}.

\begin{table}[b]
\caption{\label{table2} 
The required spacing ($\Lambda^*$ in units of $a_0$) between Ba-rich rows 
for optimal
enhancement of the DOS at $E_F$ for some dopings. The unenhanced
$N(E_F)$ is in states/Ry/Cu, and $V^*$ in mRy.}
\vskip 5mm
\begin{center}
\begin{tabular}{l c c c c c }
\hline
  $h$ & f & $N(E_F)$ & $h^*$ &  $V^*$ & $\Lambda^*$  \\
\hline \hline
 
 0.05 & 0.33 & 22 & 0.10 & 4 & 6.7    \\
 0.10 & 0.42 & 23 & 0.14 & 6 & 4.2  \\
 0.15 & 0.54 & 22 & 0.13 & 6 & 3.6  \\
 0.20 & 0.58 & 21 & 0.15 & 7 & 2.9   \\
 0.48 & 2    & 12 & -0.24 &-20 & 4.2 \\
 0.25 & 3.3  & 16 & -0.18 &-11 & 13.5  \\

\hline
\end{tabular}
\end{center}
\end{table}

\begin{figure}
\includegraphics[height=6.0cm,width=8.0cm]{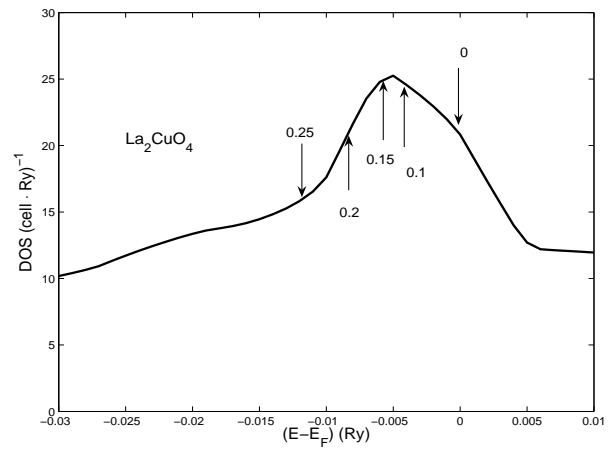}
\caption{The DOS near $E_F$ for La$_2$CuO$_4$. The arrows with the numbers indicate
the rigid-band position of $E_F$ at the respective doping (holes/Cu). For
0.5 holes $E_F$ would be at -0.030 Ry.
}
\label{fig0}
\end{figure}

The second type of superstructure is for modulations perpendicular
to the CuO-planes along the $\vec{z}$-axis.
Such modulations should be simpler
to make than in-plane modulations. For example, films with doped
and undoped layers have been made with higher $T_C$ than for optimally
doped bulk \cite{bozo2}.

Spin-polarized calculations, with applied magnetic fields of $\pm 5 mRy$ on each Cu, are made 
for 4 basic AFM unit cells above each other, with a total of
56 sites (4 CuO layers) in the cell. 
Two dopings, $h$=0.125 and $h$=0.25, are considered by replacing
1 or 2 La with Ba.  The replacement with 2 Ba surrounding a Cu layer
can be made in three different ways. One possibility ("zigzag") is when one Ba is put above
one of the two Cu 
in the first layer, while the other Ba is put below the second Cu in the 
same layer. 
Another ("column")
is when only the La close to the first Cu are replaced with Ba, 
and the third possibility ("plane") is
to fill one whole La layer with Ba. The average $h$ is 0.25, but the structures can be viewed as 3 undoped
layers sandwiched between single layers of heavily doped LBCO.
The DOS near $E_F$ show no clear gap structures, but there are variations of the local exchange
enhancements.

Table III displays the results.
The charge transferred from Cu atoms near the Ba dopant is
important, as shown by the lowest charges in Table III, and
the largest charges on Cu within La rich regions can even surpass the Cu charge
for the undoped case. 
The first four lines, for perfectly delocalized doping in the VCA,
show the general trend of weaker AFM spin waves at large doping
(Ferromagnetism, FM, appears in the VCA for much higher $h$ \cite{bba}).
The moment per Cu is highest ($\sim 0.14 \mu_B$) in the
undoped case, and it
goes down with doping to reach $\sim 0.06$ for $h$=0.375 holes per Cu.
The calculations using a supercell distribution of 1 or 2 Ba substitutions
show that larger moments (averaged and even more locally) are obtained
if the dopants are distributed within few layers only. For instance,
for 2 Ba in one plane ($h$ = 0.25) the average moment per Cu is 40 percent
larger than the moments in VCA at the same doping. The largest moments,
2.5 times larger than the VCA moments, are found within the Cu layer next to
the Ba layer. On the most distant Cu layers the moment is near 0.07,
as in VCA. By having another distribution of the two Ba near a Cu layer
("column" and "zigzag") one obtains somewhat weaker spin waves, but still
considerably stronger than seen with VCA. 

The conclusion from the result with one Ba
is partly different. The average moment is smaller than
in VCA ($0.109$), but with large local variations. The largest moments ($\sim 0.125$)
are found in the next nearest Cu layer. Moreover, this case and the 2 Ba "column"
case have another complication in common; The AFM waves on Cu are perturbed locally
near the Ba dopants so that the majority moment is considerably larger than
the minority moment in the same layer, as if FM would profit from AFM. 
Weak FM can also appear around clusters of Ba in larger supercells \cite{bba}.
In contrast,
spin-polarized calculations for AFM waves in a cell with modulated doping along $\vec{x}$
($f$=1, $\Lambda^*=8a_0$) show no sign of FM perturbations.

 \begin{table}[b]
\caption{\label{table3} 
VCA and LMTO supercell results for perpendicular modulations at different dopings.
Smallest  and largest charges ($q_{min}$ and $q_{max}$), average and maximum of
the absolute values of magnetic moments 
($\bar{m}$ and $m_{max}$) on Cu for different doping ($h$ in units of holes per Cu) for
different configurations of LBCO. All Cu-sites are equal in the results based on the VCA.
Three results for different positions of the two
La-Ba substitutions are listed ("plane", "column" and "zigzag", see the text).
The basic AFM spin configuration is imposed in all cases through application of magnetic
fields of $\pm$ 5 mRy on the Cu. The last column shows that a FM moment ($m_{FM}$, $\mu_B$
per cell) remain in two cases,
which indicates that Ba impurities perturb the imposed AFM state. }
\vskip 5mm
\begin{center}
\begin{tabular}{l c c c c c }
\hline
 $h$ & $q_{min}$ & $q_{max}$ & $|\bar{m}|$ & $|m_{max}|$ & $m_{FM}$ \\
\hline \hline
 0.000 (VCA) & 10.393 & 10.393 & 0.138 & 0.138 & - \\
 0.125 (VCA) & 10.369 & 10.369 & 0.109 & 0.109 & - \\
 0.250 (VCA) & 10.345 & 10.345 & 0.070 & 0.070 & - \\
 0.375 (VCA) & 10.322 & 10.322 & 0.063 & 0.063 & - \\
 0.125       & 10.272 & 10.455 & 0.089 & 0.125 & 0.097  \\
 0.250 "plane"& 10.207 & 10.463 & 0.098 & 0.173 & 0.000 \\
 0.250 "col." & 10.183 & 10.489 & 0.087 & 0.132 & 0.213 \\
 0.250 "zig." & 10.183 & 10.490 & 0.085 & 0.130 & 0.000 \\
\hline
\end{tabular}
\end{center}
\end{table} 

Thus, the result for modulated doping along $\vec{z}$ shows that enforcement of
spin waves can be achieved, more so locally near the doped layers than for
the average over the whole cell. This implies a larger $\lambda_{sf}$, since
$\lambda_{sf} \sim NI_m^2$, where the matrix element for spin fluctuations, $I_m$,
is proportional to $m$ \cite{tj3}, and a 
possibility for a higher superconducting $T_C$. 
It can be noted that calculations show also an enhancement of spin waves at
the surface of La$_2$CuO$_4$ \cite{tj8}. This is in line with the observation
of enhanced $T_C$ at the strained interface between undoped LBCO and SrTi$O_3$ \cite{bozo},
even though deformation and oxygen migration are important factors. Stronger effects from
combinations of parallel and perpendicular modulations seem plausible.

In conclusion, modulation of doped layers can be a possibility to enhance $T_C$ through
AFM spin fluctuations, but competition with FM might be destructive for some configurations.
Modulations along the layers seem more promising, since it can bring more states close
to $E_F$. It will be an experimental challenge to find a technique for making such modulations.
Distortion is not considered in these calculations, but periodic distortions and strain
near interfaces would be another possibility for creating potential modulations along the planes.

\end{document}